\documentclass[10pt, conference, letterpaper]{IEEEtran}

\usepackage{graphicx}
\usepackage[caption=false,font=footnotesize]{subfig}
\usepackage{caption} %
\usepackage{algorithm}
\usepackage{algorithmic}
\usepackage{amsmath}
\usepackage{url}
\usepackage{breqn}
\usepackage{cite} %
\usepackage{booktabs}
\usepackage{dblfloatfix} %
\usepackage{color}
\usepackage[per-mode=symbol,binary-units=true]{siunitx} %
\usepackage{multirow} %
\usepackage{balance}
\usepackage{lipsum}
\usepackage{hyperref} %

\IEEEoverridecommandlockouts

\begin{document}

\hyphenation{match-es}

\title{A study of HTTP/2's Server Push Performance Potential}

\author{
Rui Meireles$^\star$, Junrui Liu$^\star$, Peter Steenkiste$^\dagger$
\\ \\
\small \IEEEauthorblockA{$^\star$Computer Science Department, Vassar College, USA}
\small \IEEEauthorblockA{$^\dagger$Computer Science Department, Carnegie Mellon University, USA}
}

\maketitle

\begin{abstract}

Modern web pages have complex structures comprised of up to hundreds of different resources, such as scripts and images.
Server push is an HTTP/2 feature enabling servers to preemptively send resources to clients before they realize they need them to render a page. 
The objective is to reduce the amount of time the browser has to wait for data to be transferred, and consequently total page load time.
Our goal in this work is to quantify how much server push can actually reduce web page load times.
We approach the problem from both theoretical and experimental perspectives. We start by deriving an upper bound for the load time reduction afforded by server push. Then we proceed to actually evaluate an idealized push implementation on the Alexa Top 100 global sites\cite{alexa-top-sites}, against a non-push HTTP/2 baseline.
Our results show a linear relationship between latency and the benefit of server push. Moreover, pages with taller dependency trees tend to benefit the most from it.

\end{abstract}

\setlength{\belowdisplayskip}{10pt} \setlength{\belowdisplayshortskip}{10pt}
\setlength{\abovedisplayskip}{10pt} \setlength{\abovedisplayshortskip}{10pt}
\setlength{\textfloatsep}{10pt plus 1.0pt minus 2.0pt}
\setlength{\floatsep}{10pt plus 1.0pt minus 2.0pt}
\setlength{\intextsep}{10pt plus 1.0pt minus 2.0pt}

\section{Introduction}
\label{sec:introduction}
After its inception in the early 1990s, the web quickly became the Internet's killer application. 
Minimizing web page load time is of critical importance to improve usability and consequently user satisfaction. A 2018 study\cite{akamai-sorp} found that a \SI{2}{\second} increase in load time halves the sales conversion rate for retail sites. Another study\cite{doubleclick-tnfms} found that \SI{53}{\percent} of mobile site visits are abandoned if the page takes longer than \SI{3}{\second} to load.

Over the years, web pages have become more and more complex, a trend that is likely to continue. At present, many of the most popular pages contain hundreds of embedded objects and tens of scripts, leading to longer load times.
Given the speed of today's computers and smartphones, network transfer is most often the bottleneck in the page loading process. 

HTTP/2\cite{http2-rfc} was designed to optimize network transfer in two ways. The first is by reducing the amount of data to be transferred, which was done by changing the protocol from textual to binary and compressing the headers.
The second is by improving network transfer pipelining. Ideally, when laid out in a timeline, transfers would all be neatly concatenated without gaps. In practice this has proven difficult to achieve.

With HTTP/1.0\cite{http1.0-rfc} clients needed to wait for the reply to an outstanding request before issuing another one. This meant that for each new request, there was one Round-Trip Time (RTT) without any data flow.
HTTP/1.1\cite{http1.1-rfc} introduced pipeline support which allows multiple requests to be in-flight simultaneously. However, requests are still served in order, so a slow request at the head of the queue will block subsequent requests, introducing a \emph{pipeline bubble}. 
In practice, HTTP/1.1 browsers reduce the impact of a slow request by using multiple parallel connections. Even if one connection experiences a bubble, the other connections can still deliver objects to the browser. But using multiple connections uses more resources and increases overhead, e.g., each connection needs to go through slow start.

HTTP/2 fixes this problem by having HTTP sessions that feature multiple \emph{streams}, where each stream supports multiple pipelined HTTP requests, similar to an HTTP 1.1 session.  The server can transfer data for streams in any order, depending on what streams have data available. This solves the head-of-line blocking problem of HTTP/1.1 and obviates the need for multiple connections. The order in which requests are served is determined by a server-side scheduler, based on priorities assigned to each stream by either server or client.

HTTP/2's other transfer-optimizing feature is server push, which is the focus of our work. It consists of having servers preemptively send page-embedded objects to clients right after the page's HTML file, without waiting for an explicit request. Pushed objects are saved in the browser's cache, eliminating the RTT of latency that a network request would entail.

Push is performed by having the server send a \emph{push promise} frame announcing its intention to send the client a specific object. The client then decides to either accept (by doing nothing) or reject (by resetting the stream) the push promise. 
Rejection is useful when the client already has the object in its cache, for example.
If the stream is not reset, the server creates a new stream and sends the data after the main HTML. This can be done for potentially any page-embedded object.

While the concept of server push is simple, its practical benefits and server-side implementation are not. For example, not all objects embedded in the original main HTML are necessarily used since scripts can rewrite part of the web page.  In this paper we study HTTP/2's server push performance potential in terms of page load time reduction, both analytically and experimentally.
Our contributions are as follows:

\begin{itemize}
\item We identify and validate a theoretical upper bound for the performance potential of server push (\S\ref{sec:theoretical-analysis}).
\item We create a framework to support the execution of coherent and replicable HTTP/2 experiments (\S\ref{sec:experimental-setup}).
\item We experimentally evaluate HTTP/2's server push against a non-push HTTP/2 baseline on Alexa's Top 100 websites~\cite{alexa-top-sites} (\S\ref{sec:experimental-results}).
\item We identify the fundamental challenges in server push deployment and suggest ways to address them (\S\ref{sec:implementation-challenges}).
\end{itemize}

\begin{figure*}[t!]
\centering 
\includegraphics[width=1\textwidth]{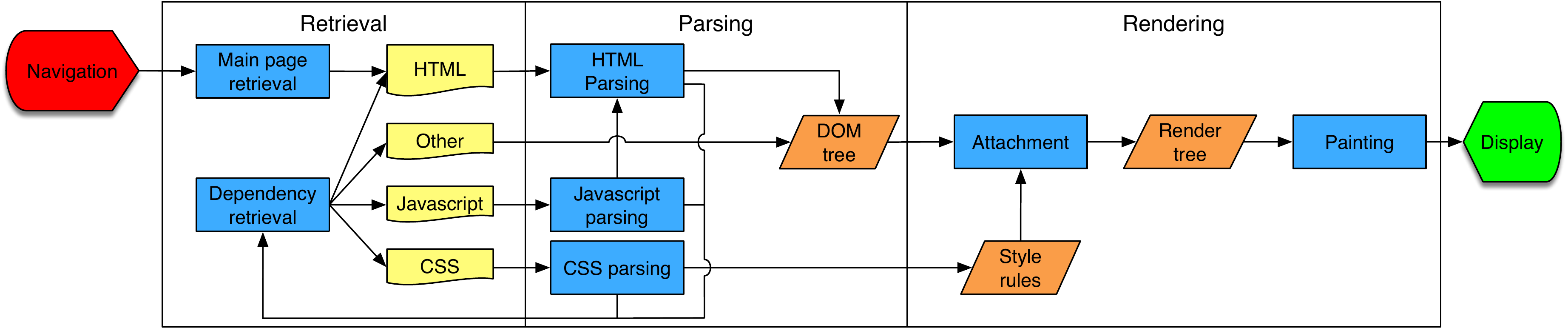}
\caption{Web page loading process.}
\label{fig:page-loading-process}
\end{figure*}

\section{Theoretical analysis}
\label{sec:theoretical-analysis}

In this section we start by analyzing how modern browsers load web pages, with the goal of understanding the sources of non-optimality. This then leads to the specification of a model for the page load time reduction potential of server push.

\subsection{Web page loading process}

An overview of the page loading process is depicted in Fig.~\ref{fig:page-loading-process}. A navigation event, for instance a link click, triggers the retrieval of the page's main HTML file. 
Parsing it yields a Document Object Model (DOM) tree describing the page's structure. As embedded objects are discovered during parsing, they are also retrieved and integrated into the DOM tree.
After parsing, formatting information from any Cascading Style Sheets (CSSs) is added to the DOM tree nodes, resulting in a render tree which is finally painted on screen. %

While both server- and client-side computations also impact the overall page load time~\cite{wang2013-demystifying}, we focus on the network transfer component of the load process. Furthermore, we assume TCP has been configured in such a way that slow start is not a performance bottleneck. We can thus define a lower bound for Page Load Time (PLT) on a cold connection as:
\begin{equation}
PLT \geq 4 \times RTT + \dfrac{psize}{bw},
\label{eq:plt-lower-bound}
\end{equation}

where $4\times RTT$ represents the TCP/TLS handshake time, in seconds, and $\frac{psize}{bw}$ represents the amount of time it takes to transfer the total amount of data needed to load the web page, $psize$ bits\footnote{For simplicity, we ignore protocol overhead when calculating $psize$.}, over a channel with bandwidth $bw$ \SI{}{\bit\per\second}.

To approach this bound, data transfers must all be neatly pipelined one after another so the network is never idle. 

HTTP/2 multiplexing aims to enable this, but is not a complete solution.  There are multiple reasons why embedded objects can introduce pipeline bubbles. %
First, the browser must start parsing the main HTML in order to discover and request the first embedded object.  Even if the browser does this as early as possible -- i.e., even before the HTML finishes downloading -- a pipeline bubble will be introduced if the RTT is longer than the HTML transfer time remaining at the time the request is issued. Since multiple requests can be issued in parallel, subsequent dependencies found while parsing the same file will not cause additional bubbles.
The same logic applies to all embedded objects requiring parsing (i.e. HTML, Javascript, and CSS files): the first dependency request may trigger a bubble.

Second, Javascript dependencies pose an additional problem. Since its execution can alter the DOM tree through calls to \texttt{document.write()}, the inclusion of an external script traditionally blocks parsing until it is retrieved from the network, delaying the discovery of any additional dependencies. Developers can mark scripts that do not change the DOM with the \texttt{async} tag to prevent blocking, but often do not. In practice, browsers avoid blocking by continuing processing on the assumption that the script will not change the DOM. If this later turns out to be false, results are discarded.

Finally, when a page embeds an object hosted on a different server, the browser will need to connect to it prior to issuing the request, introducing a delay of up to four RTTs, and potentially a pipeline bubble. This can not be avoided. Consistent with our goal of assessing server push's maximum potential, in this work we assume a whole-site serving paradigm, in which all page resources are downloaded from the same server.

\subsection{Server push performance potential upper bound}
We now derive a theoretical upper bound for the absolute Page Load Time (PLT) reduction provided by server push, which we shall refer to as SPR (for Server Push Reduction). I.e., $SPR = PLT_{no push} - PLT_{push}$, where $PLT_{no push}$ and $PLT_{push}$ represent the load times of a given page without and with push, respectively.
Throughout, we assume network transfer to be the bottleneck of the loading process.  %

\begin{figure}[b]
\centering
\includegraphics[width=0.3333\textwidth]{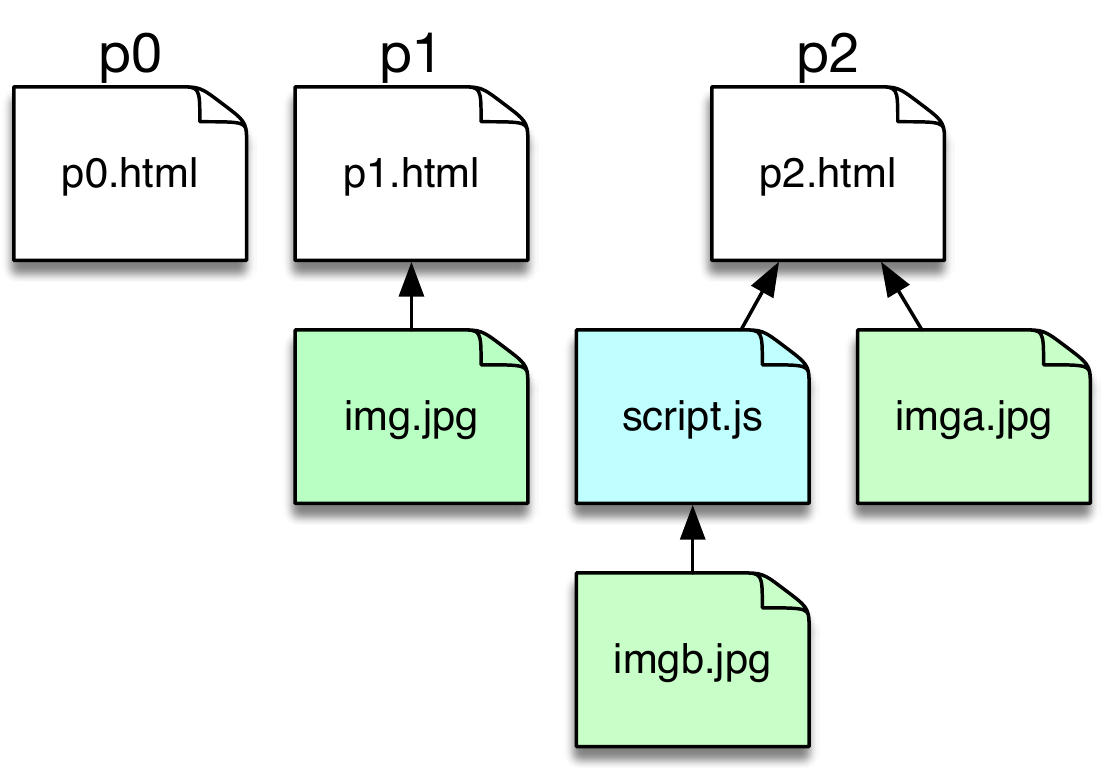}
\caption{Example web page dependency trees.}
\label{fig:example-dependency-trees}
\end{figure}

For our purposes, we can describe a web page as a dependency tree of resources. Fig.~\ref{fig:example-dependency-trees} shows three examples: 

\begin{LaTeXdescription}
	\item[\texttt{p0}:] a single, dependency-free \texttt{html} file. 
	\item[\texttt{p1}:] a single \texttt{html} file with a single image dependency. 
	\item[\texttt{p2}:] a single \texttt{html} file with two dependencies: an image and a script. The script has, itself, one image dependency.
\end{LaTeXdescription}

\begin{figure*}[b!]
\centering 
\subfloat[Web page recording]{\label{fig:exp-setup-recording}\includegraphics[width=0.352\textwidth]{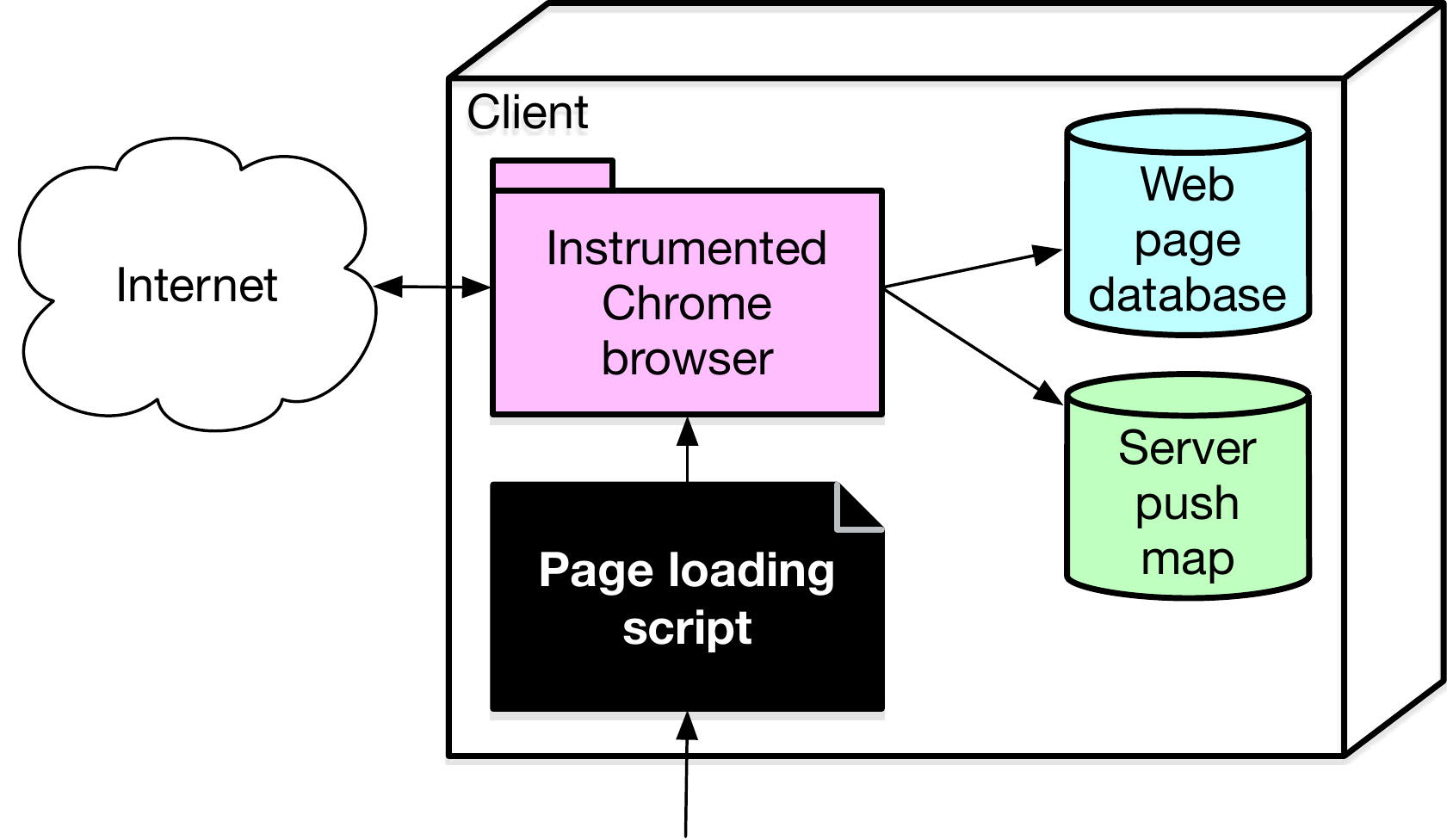}}
\hfill
\subfloat[Page load time measurement]{\label{fig:exp-setup-measuring}\includegraphics[width=0.55\textwidth]{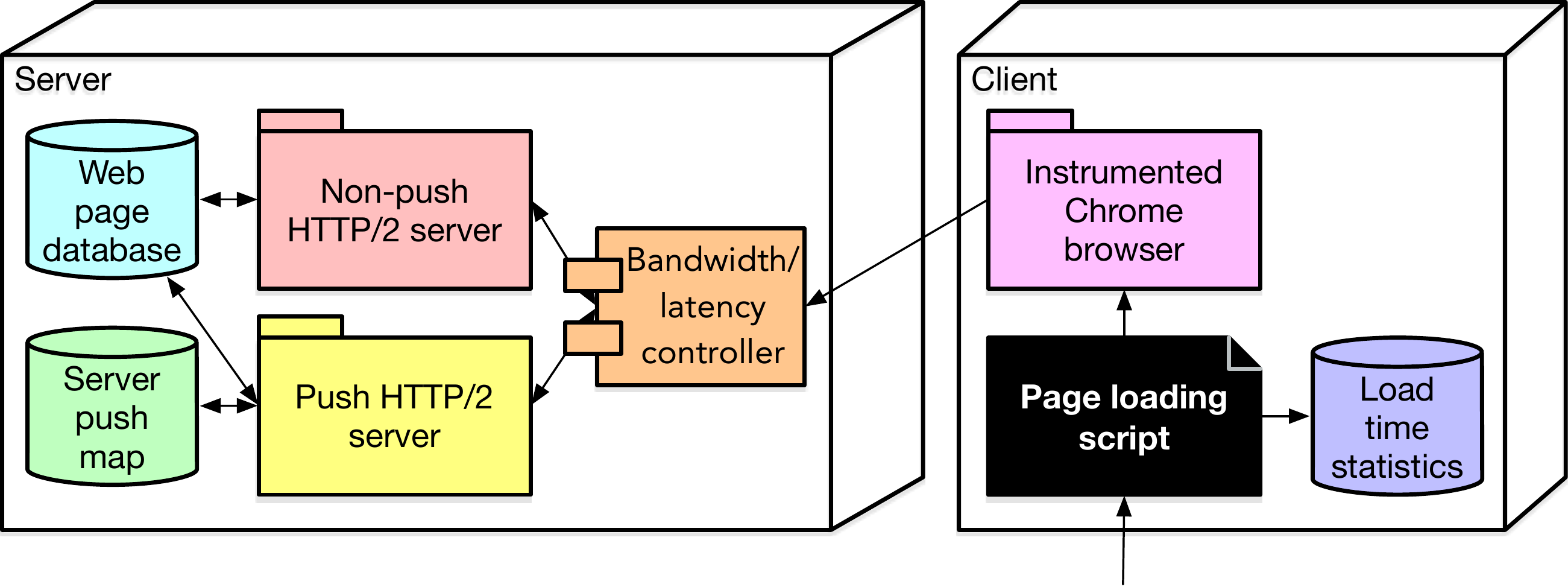}}
\caption{Experimental setup used for web page recording and for measuring page load times with and without server push.}
\label{fig:experimental-setup}
\end{figure*}

As argued in the previous subsection, for each new level in the dependency tree, a pipeline bubble of length up to RTT \SI{}{\milli\second} may be introduced. An ideal server push implementation can eliminate these bubbles entirely. Therefore, we can write an upper bound for SPR as:
\begin{equation}
SPR \leq RTT \times h,
\label{eq:spr-upper-bound}
\end{equation}
where $h$ is the height of the page's dependency tree. Using the experimental setup later described in \S\ref{sec:experimental-setup}, we validated this bound by loading the pages of Fig.~\ref{fig:example-dependency-trees} using both push and non-push versions of the same HTTP/2 server.
Minimal HTML and Javascript were used. The images were \SI{128}{\kibi\byte} each, and bandwidth was held constant at \SI{100}{\mega\bit\per\second}.

Tab.~\ref{tab:basic-spr-validation} shows the mean SPRs measured over 10 loads for various RTTs. 
Results matched expectations. Small deviations are normal since our model does not account for computation time or protocol overhead. \texttt{p0}, \texttt{p1} and \texttt{p2} all showed improvements roughly equal to RTT times their dependency trees' heights, which are 0, 1, and 2, respectively.

\begin{table}[ht!]
\centering
\begin{tabular}{|c|c|c|c|c|}
\hline
\textbf{Page/RTT} & \textbf{25 ms} & \textbf{50 ms} & \textbf{100 ms} & \textbf{250 ms} \\ \hline
\textbf{p0}            & -0.77          & -0.68           &   1.19        & -1.76            \\ \hline
\textbf{p1}            & 29.14          & 48.93          & 104.57          & 248.66          \\ \hline
\textbf{p2}            & 53.17          & 99.63         & 201.89          & 510.8          \\ \hline
\end{tabular}
\caption{Measured SPR for the pages of Fig.~\ref{fig:example-dependency-trees}}.
\label{tab:basic-spr-validation}
\end{table}

Note however, that even with an ideal push implementation capable of eliminating all pipeline bubbles, Eq.~\ref{eq:spr-upper-bound} is a loose upper bound for SPR. 
This is because the bubble introduced at a given depth $d$ in the dependency tree will be partially or completely masked if there is still data from lower-depth dependencies being transferred.
E.g., consider the discovery of \texttt{imgb.jpg} when loading \texttt{p2} from Fig.~\ref{fig:example-dependency-trees}. It is possible for this dependency to be found while \texttt{imga.jpg} is still being transferred. If the remaining transfer time for \texttt{imga.jpg} is longer than the RTT, server push will yield no practical benefit, as there will be no bubble to begin with.
We can leverage this knowledge to write a tighter upper bound for SPR:
\begin{equation}
SPR \leq \sum_{d=1}^{h} \max \{ RTT -  \dfrac{rsize_0^{d-1}}{bw}, 0\},
\label{eq:spr-approx}
\end{equation}
where $d$ represents the different depths of nodes in the dependency tree, and $rsize_0^{d-1}$ the amount of data still to be transferred for dependencies at depth levels in $[0, d-1]$ at the time of the discovery of the first dependency at depth $d$.

For each depth $d$, the gain from push can vary between 0, if there is enough data to completely mask the RTT, and RTT \SI{}{\milli\second}, if there is none.
This simple model lets us infer that server push will be most helpful in high bandwidth, high latency scenarios, as there will not be enough inflight data to mask the bubbles. Also, pages with deeper dependency trees will tend to benefit more from push.

\section{Experimental setup}
\label{sec:experimental-setup}

The goal of our experiments was to measure the potential page load time reduction from server push (SPR) on a representative set of web pages. Our strategy was to record loads for the Alexa top global 100 sites\cite{alexa-top-sites} and then reload them with server push turned on and off, noting the difference.

\subsection{Web page recording}
Fig.~\ref{fig:exp-setup-recording} depicts the setup used for recording the page loads. We employed a custom version of the Chromium browser, release 72.0.3626.112. We used it to save to disk the contents and headers of every downloaded resource. %
These were then used to populate page and server push databases for use with our HTTP/2 server.

Page load can be affected by client-side nondeterminism. For instance, pseudo-random numbers in Javascript may trigger different code paths, or be part of a URL requested from the server. In order to ensure reproducibility, we had to remove all sources of nondeterminism from the browser. This involved fixing the date, output of the pseudo-random number generator, and size of all graphical elements.

\subsection{Page load time measurement}

Fig.~\ref{fig:exp-setup-measuring} represents the setup used for performance evaluation.
We defined the metric of interest, Page Load Time (PLT), as time elapsed between the page navigation command and the firing of the browser's page load event.
Pages were served by a custom, \texttt{node.js}~\cite{node.js}-based, HTTP/2 server. It features a push control flag that can be turned on or off, so we could isolate its effect.

In order to evaluate push's maximum performance potential, we pushed all dependencies that were requested in the original page load. I.e., the main HTML was the only explicit request from the browser, everything else was pushed.

The priority with which resources are pushed is important. E.g., if images are pushed before scripts, parsing will be delayed, increasing PLT. After extensive experimentation, we decided on the following order:

\begin{enumerate}
    \item CSSes are pushed first, in the order they were originally requested. We found that since they're involved in rendering, pushing them first improves performance.
    \item All other resources follow, in the order they were requested in the original load.
\end{enumerate}

We used the Chromium browser as a client, this time modified to intercept all network requests and redirect them to our server, allowing it to impersonate the original one. Browser cache was disabled. And, in order to eliminate external factors, both server and client were deployed on the same machine.

In \S\ref{sec:theoretical-analysis} we identified transfer rate and RTT as the key parameters influencing SPR for a given web page. We then needed to select appropriate ranges for them.
The 2020 Worldwide Broadband Speed League\cite{2020-broadband-speed-league} measured mean download rates on a per-country basis.
Worldwide, the values ranged between \SI{1}{} and \SI{230}{\mega\bit\per\second}, with a mean of \SI{25}{\mega\bit\per\second}. The mean for Western Europe and North America combined was \SI{77}{\mega\bit\per\second}.
These values are expected to steadily increase over time.

\begin{figure}[b!]
\centering
\includegraphics[width=0.34\textwidth]{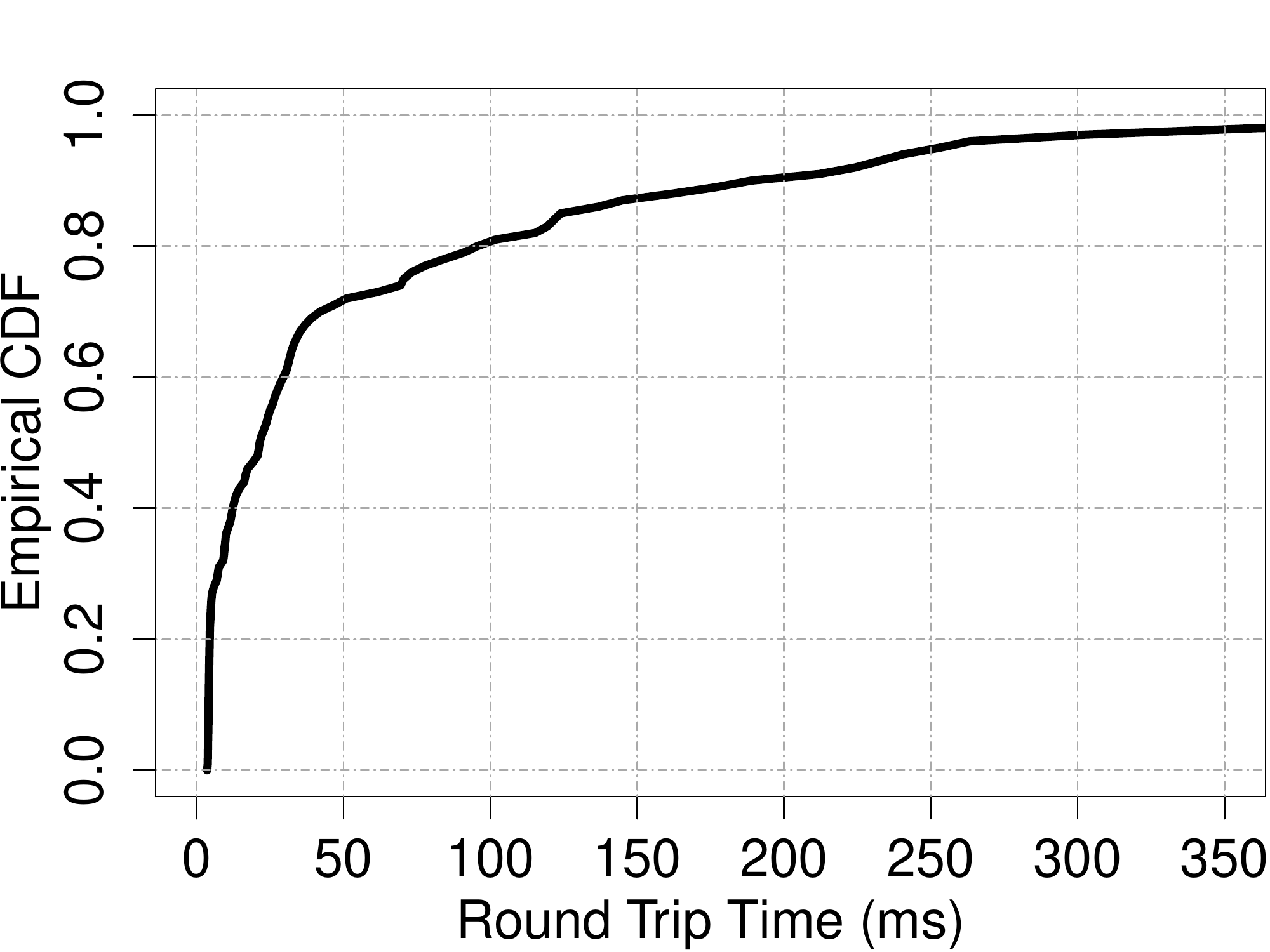}
\caption{Measured RTTs for the Alexa top global 100 sites}
\label{fig:rtt-ecdf}
\end{figure}

For RTT we could not find relevant data, so we collected some ourselves.
From our location (omitted due to double-blindness), we pinged the Alexa Top 100 sites every 15 minutes over a week-long period. 
Fig.~\ref{fig:rtt-ecdf} shows the result's Empirical Cumulative Distribution Function (ECDF). Values of \SI{5}{}-\SI{40}{\milli\second} were prevalent, and entries above \SI{250}{\milli\second} residual.

Based on this information, we picked a representative set of bandwidths (\SI{20}{}-\SI{500}{\mega\bit\per\second}) and RTTs (\SI{25}{}-\SI{200}{\milli\second}) for our experiments (server push has negligible effect on very low latency connections).
We used macOS' \texttt{dnctl} utility to control both RTT and bandwidth. For each parameter combination, we loaded each page a total of 15 times: 5 warmup loads, and 10 measured ones. Mean PLTs were computed from the latter. To ensure loads were completely cold, we restarted both browser and server between each load.

Our browser modifications and custom server created unsolvable issues, such as freezes, for a small portion of pages. Out of the 100 sites, we were able to successfully test 91.

Finally, we wanted to understand how efficient both push and non-push page loads are in absolute terms. 
To this end we created a simplified version of each web page, composed of a single HTML file whose size is equal to the size of the totality of the resources that make up the original page. The contents of this HTML were an appropriately-sized HTML comment (i.e., something of the form \texttt{<!--...-->}), so the browser did not need to render anything. We then measured the amount of time it took to load these simplified pages. Since they do not have any dependencies, their load times can be used as a lower bound for the load time of the actual pages.

\section{Experimental results}
\label{sec:experimental-results}

We analyze the effect of enabling HTTP/2 server push on Page Load Time (PLT), using the setup described in \S\ref{sec:experimental-setup}.

\begin{figure*}[t]
\centering
\subfloat[\SI{25}{\milli\second} RTT]{\label{fig:plt-ecdf-25ms}\includegraphics[width=0.45\textwidth]{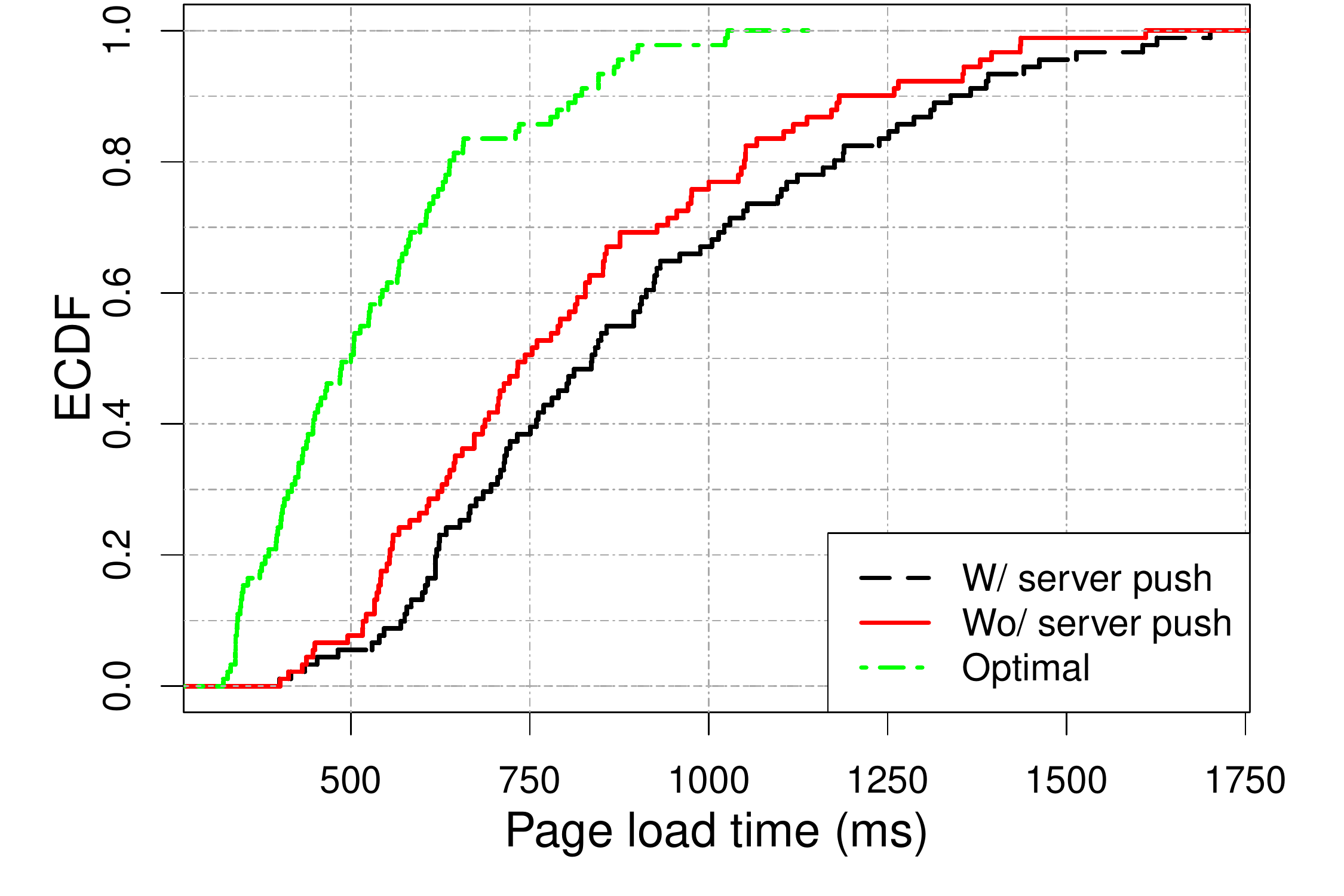}}
\subfloat[\SI{50}{\milli\second} RTT]{\label{fig:plt-ecdf-50ms}\includegraphics[width=0.45\textwidth]{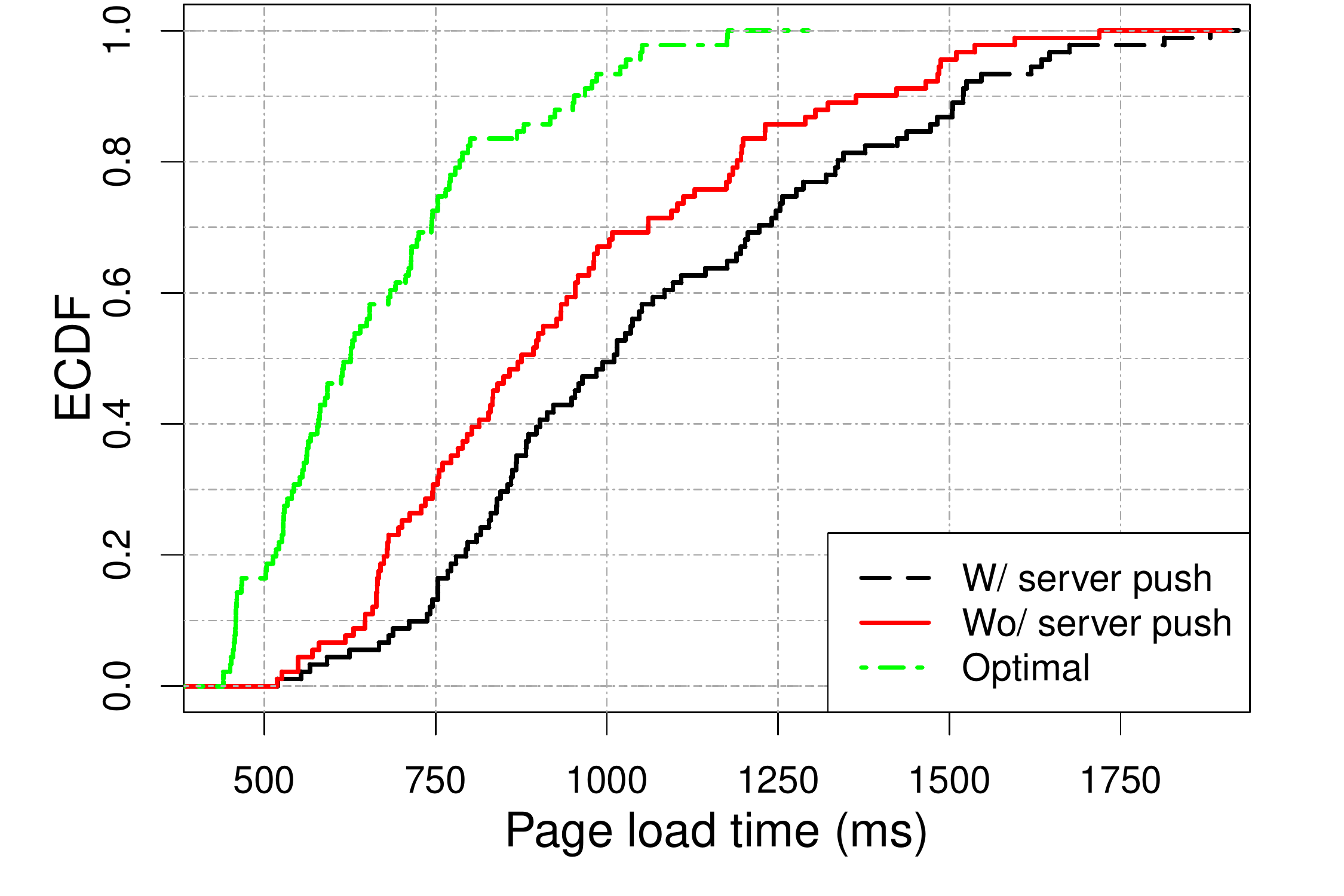}}

\subfloat[\SI{100}{\milli\second} RTT]{\label{fig:plt-ecdf-100ms}\includegraphics[width=0.45\textwidth]{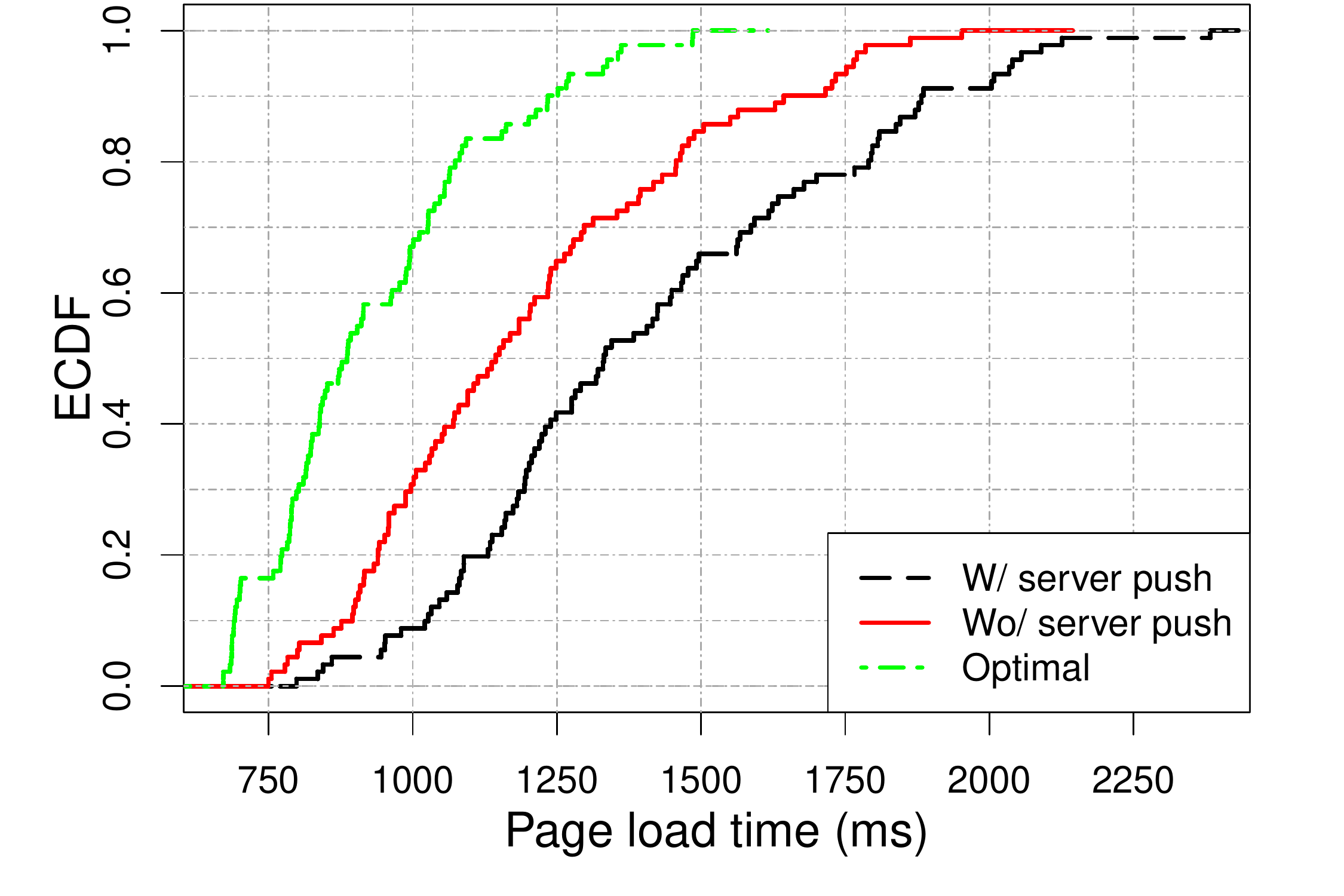}}
\subfloat[\SI{200}{\milli\second} RTT]{\label{fig:plt-ecdf-200ms}\includegraphics[width=0.45\textwidth]{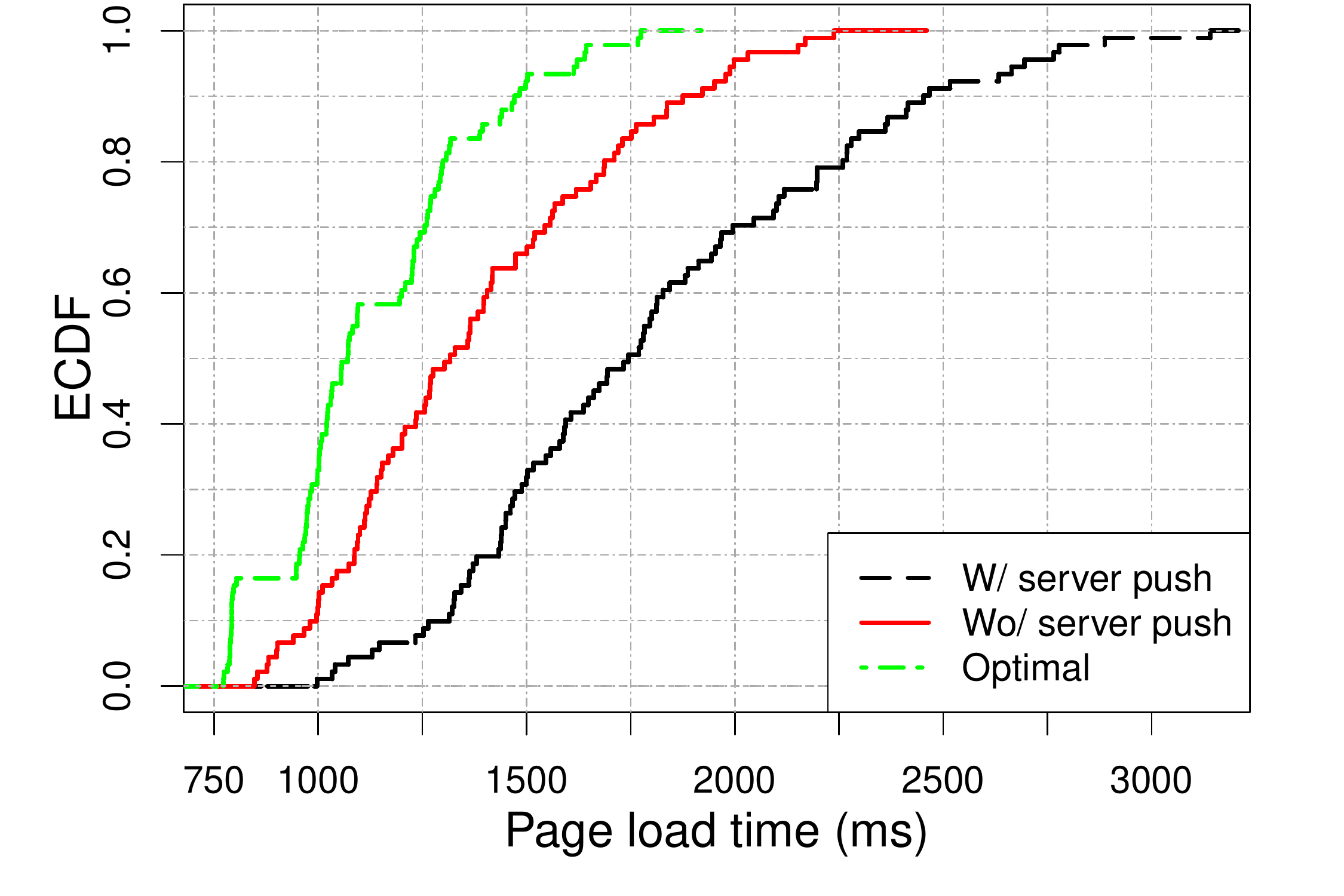}}
\caption{Page load time Empirical Cumulative Distribution Function (ECDF) for different load types: HTTP/2 without push, with push; and size-optimal. Bandwidth was held constant at \SI{100}{\mega\bit\per\second} while RTT was varied between \SI{25}{} and \SI{100}{\milli\second}.}
\label{fig:plt-ecdf}
\end{figure*}

\paragraph{\textbf{Page load time function of RTT}}

Fig.~\ref{fig:plt-ecdf} shows the ECDF of page load time across the test set, for the different types of load under consideration: HTTP/2 with and without push, and optimal. Bandwidth was held at \SI{100}{\mega\bit\per\second} while RTT varied between \SI{25}{} and \SI{100}{\milli\second}.

Two things are clear from the plots: i) enabling server push did reduce Page Load Times (PLTs), and ii) the effect was positively correlated with RTT. Both are consistent with our expectations. Server push reduces load times by eliminating pipeline bubbles where the browser is waiting for the network. And the larger the RTT, the larger the bubbles.

Server push was unable to match the load times of the optimal reference scheme that loaded a single HTML file of the correct size. This can be explained by two factors:

\begin{enumerate}

\item Load time depends not only on how long it takes to transfer the required data, but also on how long it takes to process it. Having a single resource reduces server-side processing time, as well as client-side parsing and rendering times. As the HTML contains only a comment, there is little to parse, and naught to render.

\item Downloading a single resource, instead of multiple, minimizes protocol overhead. E.g., no push promise frames need to be sent. This means fewer bytes need to be transferred overall.
\end{enumerate}

Ergo, due to its nature, the optimal scheme represents a desirable goal but is impossible to match in practice.

\begin{table}[ht!]
\centering
\begin{tabular}{|c|c|c|c|}
\hline
\multirow{2}{*}{\textbf{RTT (ms)}} & \multicolumn{2}{c|}{\textbf{Median page load time (ms)}} & \multicolumn{1}{l|}{\multirow{2}{*}{\textbf{Delta (ms)}}} \\ \cline{2-3}
                                   & \textbf{Optimal}          & \textbf{Server push}         & \multicolumn{1}{l|}{}                                     \\ \hline
25                                 & 499.8                     & 743.1                        & 243.3                                                     \\ \hline
50                                 & 625.9                     & 875.7                        & 249.8                                                     \\ \hline
100                                & 886.3                     & 1143.7                       & 257.4                                                     \\ \hline
200                                & 1071.2                    & 1317.0                         & 245.8                                                     \\ \hline
\end{tabular}
\caption{Median PLT delta between optimal and push loads.}
\label{tab:median-plt-delta-opt-vs-push}
\end{table}

\begin{figure*}[t]
\centering
\subfloat[SPR function of RTT (\SI{100}{\mega\bit\per\second} bandwidth)]{\label{fig:spr-fo-rtt}\includegraphics[width=0.4\textwidth]{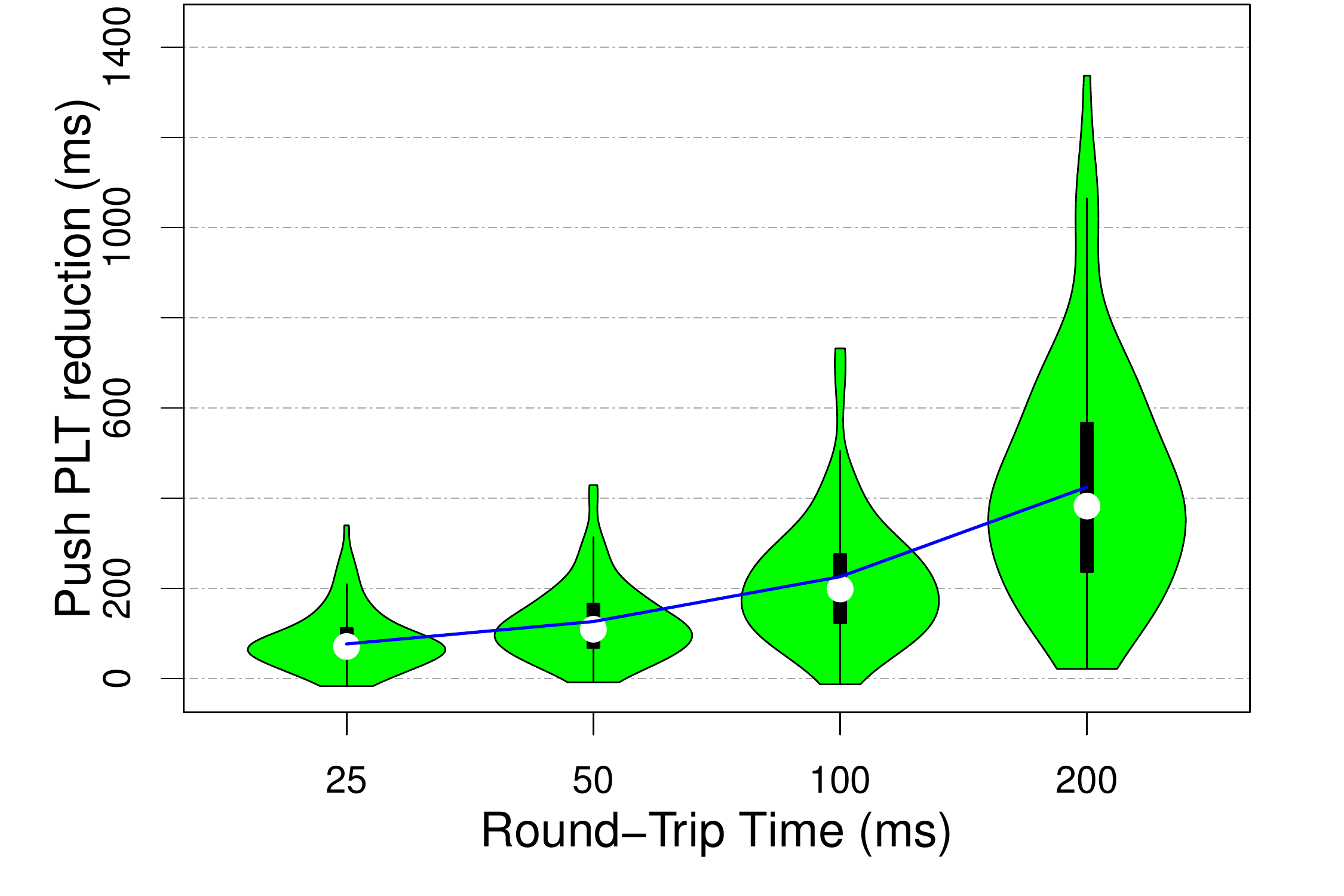}}
\subfloat[SPR function of bandwidth (\SI{200}{\milli\second} RTT)]{\label{fig:spr-fo-bw}\includegraphics[width=0.4\textwidth]{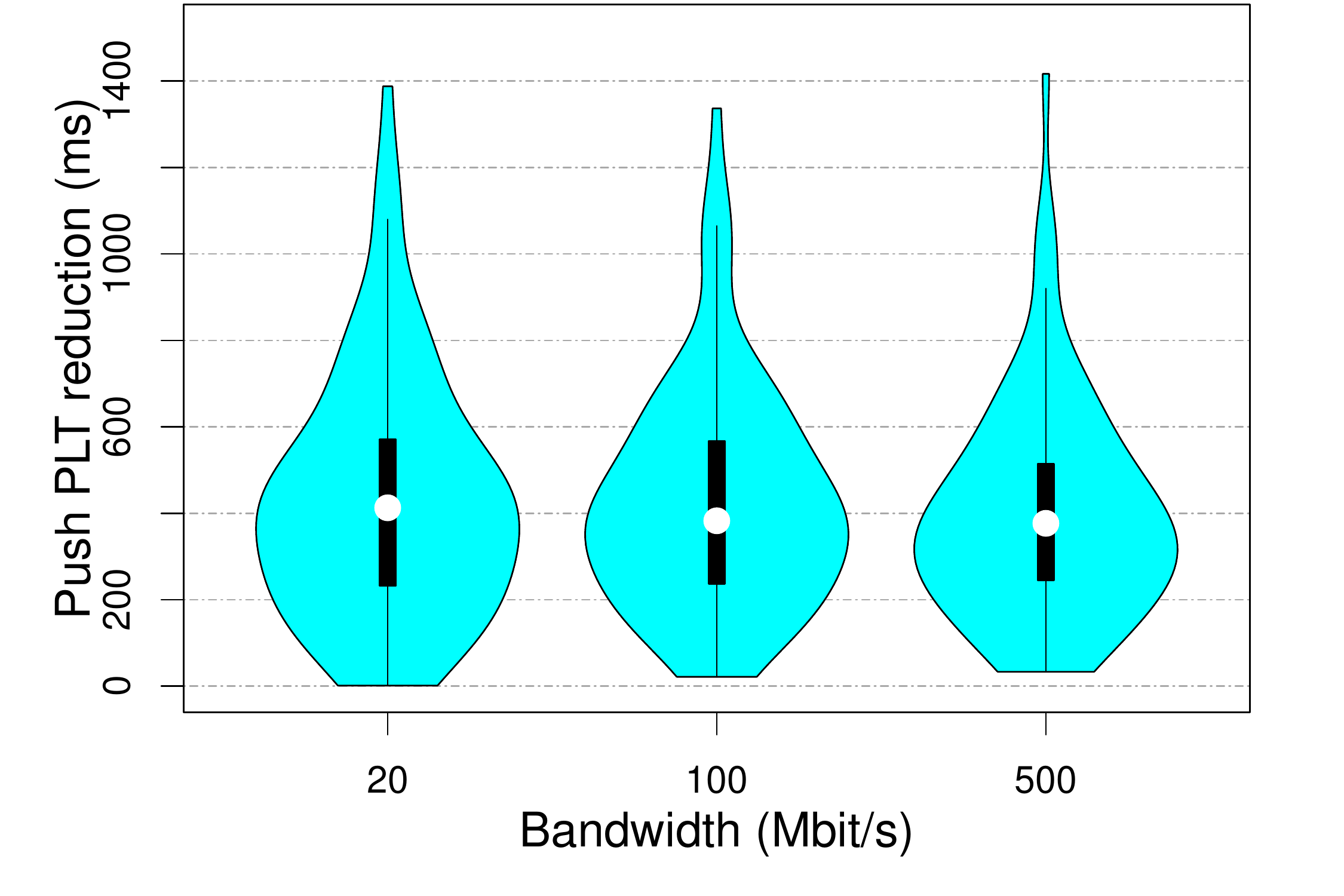}}
\caption{Impact of the RTT and bandwidth parameters on the load time reduction achieved by enabling server push (SPR).}
\label{fig:spr-parameter-space}
\end{figure*}

It is interesting to note that the delta between the optimal and server push schemes appears to be almost independent of RTT, as Tab.~\ref{tab:median-plt-delta-opt-vs-push} demonstrates for median deltas.

The fact that both schemes remained separated by around \SI{250}{\milli\second} across RTTs indicates that server push successfully eliminated all pipeline bubbles. If there were bubbles left, they would grow larger with RTT, increasing the load time delta. 
The difference between push and optimal loads must therefore be attributed to different processing times and protocol overheads, as described earlier.

\paragraph{\textbf{Server Push time Reduction (SPR) function of RTT}}

Let us now focus on the relationship between the load time reduction afforded by server push (SPR) and RTT. From the analysis of \S\ref{sec:theoretical-analysis} we anticipate SPR to grow linearly with RTT. Fig.~\ref{fig:spr-fo-rtt}, which shows SPR violin plots for different RTTs (bandwidth was fixed at \SI{100}{\mega\bit\per\second}), confirms this bore out in practice. 
Each violin displays the SPR median, quartiles, and probability density. 
We can see that doubling the RTT roughly doubled the median SPR, e.g., when RTT goes from \SI{100}{} to \SI{200}{\milli\second}, median SPR goes from $\sim$\SI{200}{} to $\sim$\SI{400}{\milli\second}. 

A least mean-squares linear regression on SPR as function of RTT yields $SPR = 1.99 RTT + 27.16$, which is shown as a blue line in Fig.~\ref{fig:spr-fo-rtt}. Using this equation for SPR prediction yields a Median Absolute Deviation (MAD) of \SI{65.65}{\milli\second}, which is reasonable given that page structure also affects SPR.

Our theoretical analysis also determined that the product of RTT and dependency tree height forms an SPR upper bound (Eq.~\ref{eq:spr-upper-bound}). The mean dependency tree height of the tested pages was \SI{2.7}{}, and the median \SI{2}{}. The \SI{1.99}{} regression coefficient is consistent with these values.

\begin{figure}[t]
\centering
\includegraphics[width=0.37\textwidth]{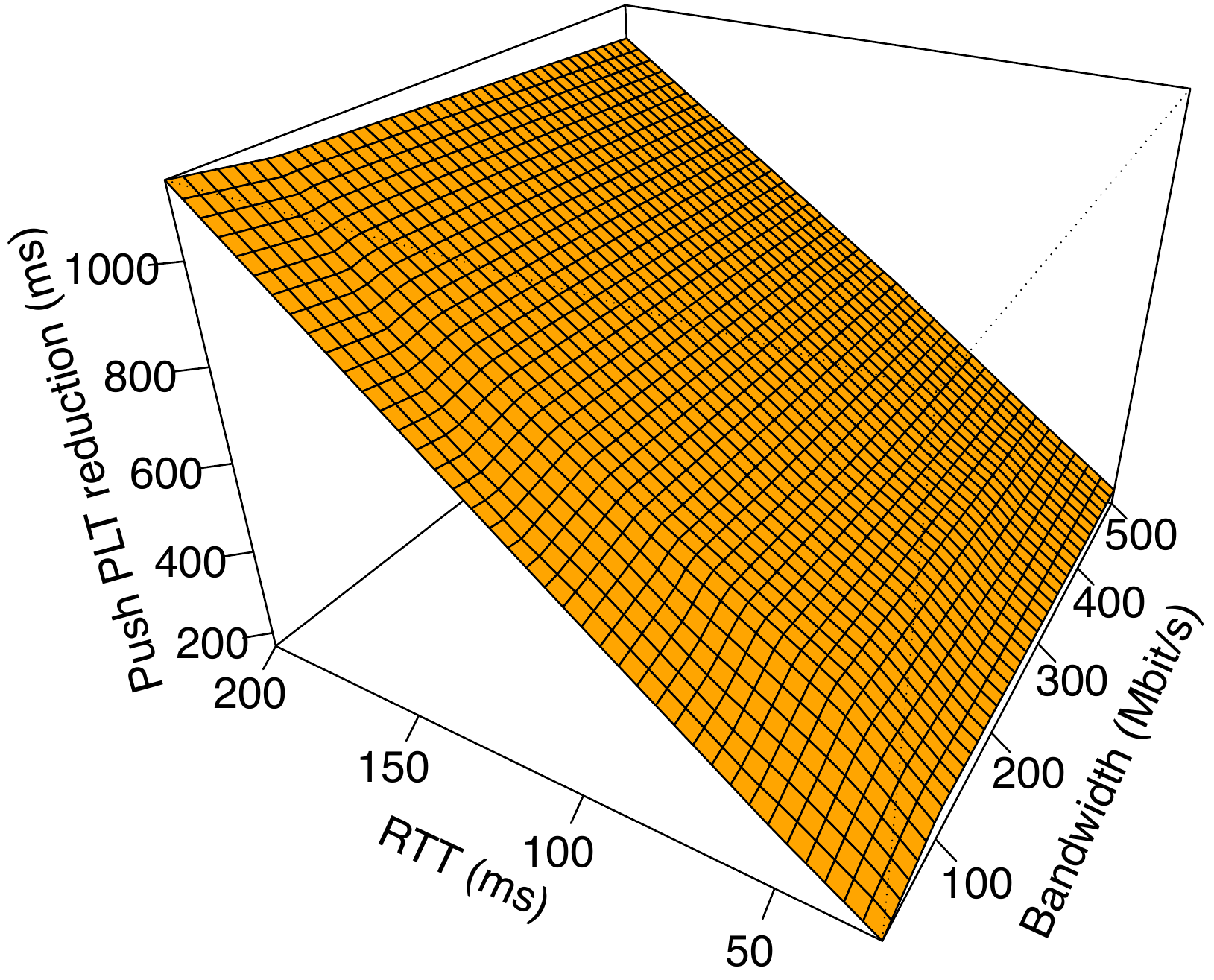}
\caption{SPR for BBC.com (dep. tree height 6).}
\label{fig:spr-bbc.com}
\end{figure}

\paragraph{\textbf{SPR function of bandwidth}}
Eq.~\ref{eq:spr-approx} sets our expectations for the effect of bandwidth on SPR. If, when a new request is issued, data from prior requests is still pending, the latency associated with the new request will be at least partly hidden. Since lower bandwidth yields slower transfers, we anticipated that if we decreased bandwidth, SPR would decrease as well. In practice, that was not the case.

Fig.~\ref{fig:spr-fo-bw} shows SPR violin plots for different bandwidths, with RTT being held at \SI{200}{\milli\second}. SPR appears to be almost unchanged across the range of tested bandwidths.
From studying the page loading logs, for higher bandwidths server push isn't able to fill in all the RTT gaps because the TCP window is not large enough on a cold connection. So loading halts waiting for TCP to catch up. Further TCP parameter tuning or usage of warm connections, whose windows are already large, would be necessary to overcome this issue.

Fig.~\ref{fig:spr-bbc.com} shows SPR plotted against both RTT and bandwidth for \texttt{bbc.com}'s homepage. A clear linear relationship between RTT and SPR can be seen. In contrast, the effect of bandwidth is minor. Both are representative of our overall results.

\paragraph{\textbf{SPR function of dependency tree height}}
Finally, we investigate the impact of page structure on SPR. Our analysis identified resource dependency tree height as the key structural factor. In most instances, Chromium provides information about which resource triggered a given network request. We used this information to reconstruct each page's resource dependency tree as well as possible.

\begin{figure}[t]
\centering
\includegraphics[width=0.42\textwidth]{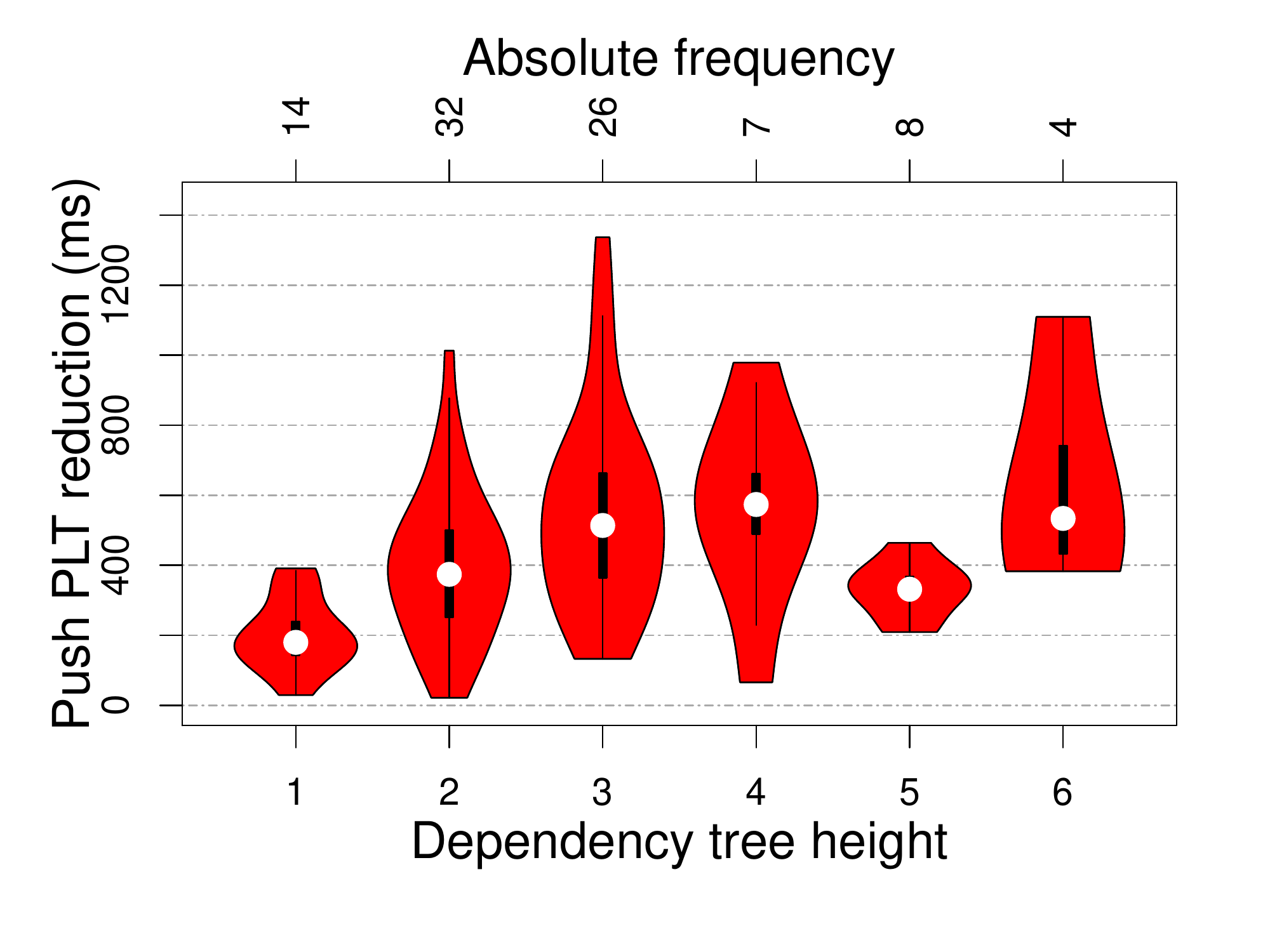}
\caption{SPR function of dependency tree height (\SI{100}{\mega\bit\per\second}, \SI{200}{\milli\second} RTT).}
\label{fig:spr-fo-dtreeheight}
\end{figure}

Fig.~\ref{fig:spr-fo-dtreeheight} depicts SPR as a function of dependency tree height for one of our experiments (\SI{100}{\mega\bit\per\second} bandwidth, \SI{200}{\milli\second} RTT), and is representative of the overall results. Positive correlation between tree height and SPR can be observed, which matches our expectations. 
In particular, for the subset of pages with dependency trees of heights between 1 and 3, which represents \SI{79}{\percent} of all tested pages, there is a clear linear relationship between tree height and SPR. This is not the case, however, for pages with taller trees. This can be explained in part by the fact that the sample size for taller tree heights was smaller. Because of this, other page characteristics such as parsing and rendering times may have skewed the results.

Also interesting to note is the fact that, for some pages, SPR exceeded the theoretical limit of Eq.~\ref{eq:spr-upper-bound}: dependency tree height times RTT. This is because Eq.~\ref{eq:spr-upper-bound} does not take processing time into account. In practice, finishing resources early means parsing can start earlier as well, and execute with fewer interruptions, leading to additional time gains, e.g., from less context switching. Furthermore, our push implementation prioritized CSS resources, allowing rendering to start early.

\section{Implementation challenges}
\label{sec:implementation-challenges}

In our performance evaluation we used pages that always loaded in the same exact fashion, and were therefore able to precompute a list of resources to push for each page.
In this section we describe the major hurdles to transitioning from that idealized scenario to a practical server push implementation.

\subsection{Nondeterministic requests}
The feasibility of server push hinges on the server being able to predict future client requests.
If requests are nondeterministic, that becomes impossible.

We have identified four major sources of nondeterminism, all stemming from the fact that requested URLs can include the following types of data, obtained from the Javascript API:

\begin{LaTeXdescription}

	\item[Pseudo-random values:] Values generated by calls to the \texttt{Math.random()} function.

	\item[Timing data:] Either the client's clock value, provided by \texttt{Date.now()}, or page load timing information, provided by the \texttt{window.performance} object.

	\item[Browser properties:] Values such as window size and position, browsing history length, local storage, etc. Provided by the \texttt{window} and \texttt{document} objects.

	\item[Device properties:] Values such as screen size, from the \texttt{window.screen} object, and geographical location, from the \texttt{navigator.geolocation} object.

\end{LaTeXdescription}

We believe the best the server can do is to avoid pushing any nondeterministic resources. This could be accomplished for instance by parsing the page before serving it, looking for any URLs including nondeterministic values. It would however introduce additional complexity and delay.

\subsection{Dynamic content}

Although the web started as an information archive of largely immutable documents, over time it became more and more dynamic. Today, web page content is extremely fluid.
Refreshing a Facebook feed, for instance, will yield a different page every time. Also, content will differ depending on the user's identity, each user seeing a different feed, even if both are loaded simultaneously. Finally, content can also vary according to the device used to access it, as a way to deal with heterogeneous screen and computation capabilities. %
Moreover, with the current trend towards an \emph{Internet Of Things}, the need for device-based customization is likely to increase.

In summary, modern web pages are time-, user- and device-sensitive. Ergo, there is no single list of resources to push for each web page. Rather, each page load is unique.

Potential strategies include:

\begin{itemize}

    \item Before serving it, the server can parse the page to find dependent resources to push. However, this introduces delay, partially offsetting the potential benefit. Kansal~\cite{kansal2019-alohamora} has proposed the use of a neural network for this. After training, the network chooses what to push on the fly, using page structure and client browser attributes as input.

    \item Dynamic pages are generated by server-side programs. The code that generates the page could, as it builds it, take note of the resources being referenced and communicate them to the HTTP/2 server, an idea similar in spirit to that of Netravali et al. \cite{netravali2016-polaris}. The downside is that this requires support from the page-generating application.

\end{itemize}

\subsection{Cached content}

Browsers maintain content caches. Cached content is redundant and should not be pushed. There are multiple ways to achieve this. One way to do this is to have the browser reject the push promises for resources that it has in its cache. However, some bandwidth will still be spent on the push promises and their replies.
An alternative is for the browser to send a bloom filter~\cite{bloom1970} of its cache contents as part of a cookie in the initial page request. A bloom filter is a compact data structure that enables fast set membership testing. Crucially, it is free from false negatives, meaning no cached content will be pushed. It does however, have a non-zero false positive probability, meaning non-cached resources may end up not being pushed. The filter should be made large enough to limit the false positive probability to a reasonable number.

\subsection{External resources}
Currently, many web pages include resources that are not hosted by the server serving the page's HTML. Ads and social networking plugins are examples of resources that are often obtained from external servers. This limits server push because since they are external, they can not be pushed.

An option would be to have external servers push the resources they are responsible for. But this would be difficult to implement, as the external server would need to know which resources go together without having access to the original main HTML file. One may think of creating a mapping, and having the client send a token key to index said mapping, as part of a cookie. But since many pages are generated on the fly, this would require tight coupling between the different servers, which would defeat the purpose of external resources.

An alternative is to simply not push these resources. This would reduce performance but, since they tend to be served by Content Delivery Networks (CDNs), which locate servers near clients to reduce RTT, the penalty should be small.

Also, there is interest in having CDNs transition towards whole-site serving, which would obviate this limitation. %

\section{Related work}
\label{sec:related-work}

We consider work in three areas related to page load optimization: HTTP/2 server push specifically, web prefetching at large, and dependency analysis.

\paragraph{\textbf{HTTP/2 server push}}

There have been multiple studies on the performance of HTTP/2 server push.

Wang \emph{et al.}~\cite{wang2014speedy} briefly studied push as part of SPDY~\cite{belshe2012spdy}, the protocol that eventually developed into HTTP/2. Their results showed push performs best when every resource is pushed and when RTT is high.
Rosen \emph{et al.}~\cite{Rosen:2017:PRI:3038912.3052574} also experimentally evaluated the performance potential of server push. Their results agree with ours, with push performing best on high latency and/or lossy scenarios.

Zimmermann \emph{et al.}~\cite{Zimmermann:2018:WRH:3281411.3281434} analyzed the impact of server push strategies, i.e., what resources to push and in what order, on page load time. They discovered optimal performance requires custom per-site strategies. Pushing every resource actually reduced performance for some pages. 
This is further evidence of the difficulty of designing practical push implementations, leading to low market penetration. In 2017~\cite{Zimmermann2017HowHP}, the same authors surveyed over 5 million HTTP/2 sites and measured push adoption at only around \SI{0.01}{\percent}. In 2019, adoption rate within the Alexa top 10,000 sites was reportedly \SI{0.9}{\percent}~\cite{kansal2019-alohamora}.

By providing not only experimental results but also a theoretical framework to help better understand push behavior, our work complements all these.

\paragraph{\textbf{Web prefetching}}
Prefetching is a technique where the client predicts what content the
user will need in the future, and prefetches it before the user
actually requests it. 
Prefetching web content has a long history. Some
studies~\cite{fan1999web} have focused on reducing latency at the cost
of more traffic, caused by inaccurate predictions of future content needs.
Other designs~\cite{klemm1999webcompanion} provide more modest
performance gains while minimizing the adverse effects of prediction error.

In summary, both prefetching and push try to perform transfers in
advance, to improve performance. The key difference is that prefetching is controlled by the client, while push is controlled by the server. While prefetching has the advantage of knowing client state (e.g., cache
contents), server push can typically be more effective since it knows all files associated with a page and the dependencies
between them. Both techniques come with the danger of wasted
bandwidth due to bad predictions, however.

\paragraph{\textbf{Dependency analysis}}
Modern web pages feature complex dependency graphs, hence a lot of effort has been dedicated to optimizing the load of dependent resources.

Netravali \emph{et al.}~\cite{netravali2016-polaris} proposed profiling web pages
offline and embedding the dependency graph into the page's
main HTML. A dynamic client-side scheduler then loads the objects in
optimal order with respect to the dependency graph.

In Vroom~\cite{ruamviboonsuk2017-vroom} the server also creates a dependency graph. But then, it uses server push for local dependencies, and hints of URLs for ones that need to be downloaded from other servers. This improves performance when resources from multiple servers are needed, as the client can fetch the hinted URLs before discovering them during parsing.

Other works~\cite{wang2016-speeding, netravali2019-watchtower} have advanced a split-browser approach, introducing a powerful proxy between client and server. The proxy quickly downloads the page by using its greater resources relative to the client. It then identifies the needed resources, and preemptively sends them to the client.

\section{Conclusions}
\label{sec:conclusions}
We studied the potential for HTTP/2 server push to reduce web page load times.
Our theoretical analysis uncovered a linear relationship between the potential benefit of server push and RTT, as well as a positive correlation with the height of the page's resource dependency tree.

Our experimental evaluation using an idealized setup confirmed our analytical findings. Having the server push every resource brought load times closer to the optimum, with gains growing linearly with RTT. Bandwidth, on the other hand, had little effect on results.

We conclude server push is most appealing for structurally-complex web pages delivered over high-latency connections. 

However, a practical implementation of push has major hurdles to overcome, especially the nondeterministic nature of client requests and the dynamic nature of modern web pages. Server push adoption has been slow~\cite{Zimmermann2017HowHP,kansal2019-alohamora}, and we suspect that will remain the case in the foreseeable future.

\bibliographystyle{IEEEtran}
\bibliography{IEEEabrv,pushpaper}

\end{document}